\documentclass[aps,pra,superscriptaddress,twocolumn,amsfonts,amsmath,amssymb,floatfix]{revtex4-2}
\usepackage[T1]{fontenc}
\usepackage[utf8]{inputenc}
\usepackage{graphicx}
\usepackage{xcolor}
\usepackage{bm}
\usepackage{microtype}
\usepackage[colorlinks=true,pdfstartview=FitV,linkcolor=blue,citecolor=blue,urlcolor=blue]{hyperref}
\begin{document}
\title{Interfacial Faraday instability driven by interspecies-interaction modulation in a phase-separated binary Bose--Einstein condensate}
\author{Yating Huang}
\affiliation{Department of Physics, School of Physics and Electronic Science, East China Normal University, Shanghai 200241, China}
\author{Yan Li}
\email{yli@phy.ecnu.edu.cn}
\affiliation{Department of Physics, School of Physics and Electronic Science, East China Normal University, Shanghai 200241, China}

\begin{abstract}

We investigate interfacial Faraday instabilities in a quasi-two-dimensional, phase-separated binary Bose--Einstein condensate driven by periodic modulation of the interspecies scattering length. In the strongly immiscible regime, the interspecies interaction energy is concentrated near the internal interface, linking the modulation to interfacial tension and the capillary restoring force. We quantify the first-order tension response through its logarithmic sensitivity and account for nonlinear tension variations at finite modulation amplitudes. A sharp-interface capillary model and Floquet analysis yield discrete azimuthal-mode frequencies and primary instability tongues that shift toward higher driving frequencies with increasing mode order. Gross--Pitaevskii simulations reveal star-shaped interface patterns with threefold to eightfold rotational symmetry. The eightfold mode exhibits subharmonic oscillations and approximately exponential growth of its period-averaged amplitude over a small-deformation interval. These results provide a route to frequency-selective excitation of interfacial capillary modes through interspecies-interaction control.

\end{abstract}

\maketitle
\section{Introduction}
\label{sec:introduction}

Periodic forcing can destabilize a fluid surface and excite subharmonic standing waves, as in the Faraday instability~\cite{Faraday1831,Benjamin1954,Binks1997}. At the primary parametric resonance, the response oscillates at half the driving frequency, with resonant wavelengths determined by the dispersion relation. For a closed, approximately circular boundary, the allowed azimuthal modes are discrete, so parametric excitation can produce shape oscillations with distinct rotational symmetries~\cite{Noblin2005,Shen2010,Maity2020,Kwon2021}.

Atomic Bose--Einstein condensates (BECs) allow parametric excitation through the control of interatomic interactions and confinement~\cite{Nicolin2007}. Periodic modulation of the $s$-wave scattering length was predicted to generate subharmonic density patterns~\cite{Staliunas2002}, and longitudinal Faraday waves were observed by modulating the transverse confinement of an elongated condensate~\cite{Engels2007}. Interaction-modulated condensates also exhibit driven density patterns and more complex nonequilibrium responses~\cite{Nguyen2019,Zhang2020}. Binary superfluids support density- and spin-wave Faraday responses under Rabi coupling, confinement modulation, and interaction modulation~\cite{Chen2019,Cominotti2022,Wang2025}. In a strongly oblate condensate, interaction modulation also excites star-shaped surface modes whose resonance frequencies resolve the discrete collective-excitation spectrum~\cite{Kwon2021}. Related light-induced spatiotemporal density patterns have also been observed in a superfluid Fermi gas under spatially structured optical excitation~\cite{Li2021Fermi}.

Phase separation introduces an internal interface whose shape oscillations involve both condensates~\cite{Ho1996,Pu1998,Timmermans1998,Hall1998}. Interaction-controlled miscibility has been demonstrated in $^{85}\mathrm{Rb}$--$^{87}\mathrm{Rb}$ mixtures~\cite{Papp2008}; in a trap, the density distribution also depends on the atom numbers of the two components and the trapping geometry~\cite{Lee2016}. Control of phase separation by an intercomponent Feshbach resonance has also been demonstrated in two hyperfine states of $^{87}\mathrm{Rb}$~\cite{Tojo2010}.
\textcolor{black}{In a quasi-two-dimensional (quasi-2D) geometry, radial phase separation can yield a central disk of one component surrounded by an annulus of the other, with a nearly circular internal interface}~\cite{Maity2020}.
 Interfacial tension supplies the capillary restoring force, while the motion of both condensates contributes to the mode inertia, together setting the natural frequency scale of interfacial oscillations~\cite{Mazets2002,Indekeu2018,Kobyakov2012}. Other interface instabilities in binary condensates arise from acceleration or relative flow, including Rayleigh--Taylor and Kelvin--Helmholtz instabilities~\cite{Sasaki2009,Takeuchi2010,Suzuki2010}.

\textcolor{black}{Maity \emph{et al.} investigated star-shaped interface patterns driven by periodic modulation of the core-component intraspecies interaction~\cite{Maity2020}. In their protocol, the modulation changes the bulk mean-field interaction energy and produces a time-dependent pressure imbalance across the interface. In contrast, modulation of the interspecies interaction acts through the overlap energy, which is concentrated near the internal interface in the strongly phase-separated regime~\cite{Ao1998,Barankov2002}. This distinction motivates a quantitative description of the tension-mediated capillary drive.}

\textcolor{black}{The central question is how a prescribed modulation of the interspecies scattering length changes the interfacial tension and, in turn, the capillary restoring force. At finite modulation amplitudes, the nonlinear tension response can also shift the cycle-averaged restoring coefficient and introduce higher temporal harmonics. Related control of Faraday-wave patterns through surface tension has been demonstrated in a classical liquid-metal system~\cite{Zhao2018}.}

\textcolor{black}{To address this question, we investigate the parametric excitation of azimuthal capillary modes at the internal interface of a phase-separated quasi-2D $^{87}\mathrm{Rb}$--$^{85}\mathrm{Rb}$ BEC. The equilibrium disk--annulus configuration is illustrated in Fig.~\ref{fig:equilibrium}. We quantify the interfacial-tension response through its logarithmic sensitivity and retain its nonlinear interaction dependence to construct the periodic capillary restoring coefficient, as illustrated in Fig.~\ref{fig:tension_response}. A sharp-interface capillary model and Floquet analysis then determine the discrete azimuthal-mode frequencies and primary instability tongues for $m=3$--$8$. GP simulations reveal the corresponding star-shaped interface patterns, while free and driven eightfold-mode dynamics establish the natural-frequency scale, subharmonic motion, and finite-time amplification.}

\section{Equilibrium interface and interaction modulation}
\label{sec:equilibrium_driving}

\subsection{Quasi-2D model and equilibrium interface}
\label{subsec:II_model}

The system consists of $^{87}\mathrm{Rb}$ and $^{85}\mathrm{Rb}$ condensates, denoted by components $A$ and $B$, respectively, in a strongly oblate, axisymmetric harmonic trap. The radial and axial trapping frequencies are $\omega_\perp/(2\pi)=15\,\mathrm{Hz}$ and $\omega_z/(2\pi)=1500\,\mathrm{Hz}$, respectively.

\begin{figure}[!t]
  \centering
  \includegraphics[width=0.99\columnwidth]{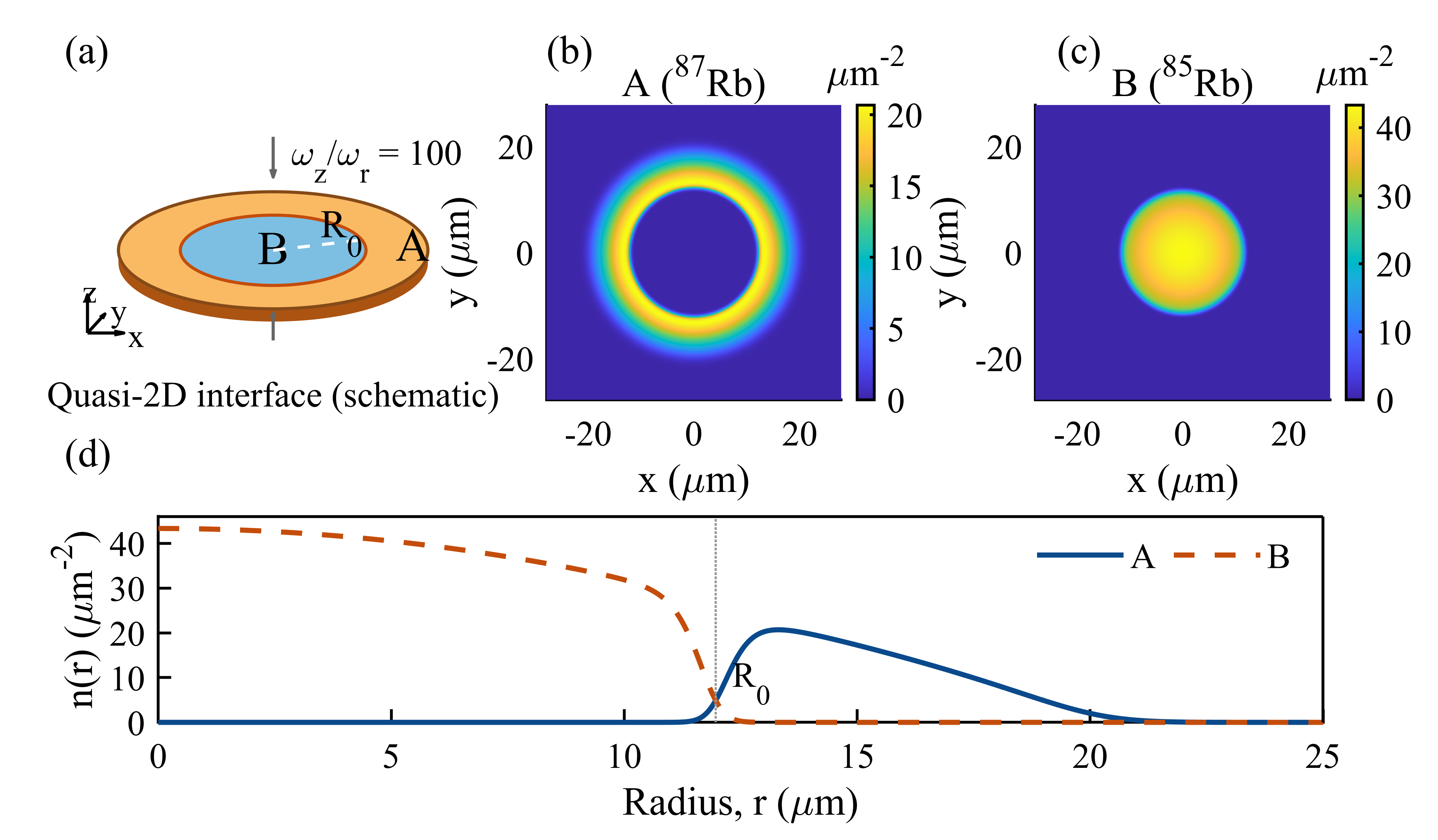}
 
\caption{Equilibrium interface and density distributions in a phase-separated quasi-2D binary BEC.
(a) Schematic of the disk--annulus configuration, with component $B$ occupying the central disk and component $A$ the surrounding annulus. Their common boundary is the internal $A$--$B$ interface.
(b,c) Equilibrium 2D density distributions of components $A$ ($^{87}\mathrm{Rb}$) and $B$ ($^{85}\mathrm{Rb}$), respectively, obtained from the stationary GP equations.
(d) Radial density profiles $n_A(r)$ (blue solid line) and $n_B(r)$ (orange dashed line). The vertical dotted line marks the equilibrium interface radius $R_0$.}
\label{fig:equilibrium}
\end{figure}

\textcolor{black}{The quasi-2D disk--annulus configuration and the internal interface separating the two components are illustrated schematically in Fig.~\ref{fig:equilibrium}(a).} We adopt a quasi-2D description with the axial motion frozen to the harmonic-oscillator ground state. The stationary condensate order parameters satisfy the coupled 2D GP equations

{\color{black}
\begin{equation}
\begin{aligned}
\mu_A\Psi_A
&=\biggl[-\frac{\hbar^2}{2m_A}\nabla_\perp^2
+V_A(\boldsymbol{\rho})\\
&\quad+g_{AA}|\Psi_A|^2
+g_{AB}|\Psi_B|^2\biggr]\Psi_A,\\[2pt]
\mu_B\Psi_B
&=\biggl[-\frac{\hbar^2}{2m_B}\nabla_\perp^2
+V_B(\boldsymbol{\rho})\\
&\quad+g_{BB}|\Psi_B|^2
+g_{AB}|\Psi_A|^2\biggr]\Psi_B.
\end{aligned}
\label{eq:II_GP}
\end{equation}
}

\textcolor{black}{Here $\Psi_j(\boldsymbol{\rho})$ denotes the stationary order parameter of component $j=A,B$, with atomic mass $m_j$ and chemical potential $\mu_j$. The in-plane coordinate is $\boldsymbol{\rho}=(x,y)$, and $V_j(\boldsymbol{\rho})=m_j\omega_\perp^2\rho^2/2$ is the harmonic trapping potential. The order parameters satisfy $\int d^2\boldsymbol{\rho}\,|\Psi_j|^2=N_j$, where $N_j$ is the atom number and $n_j(\boldsymbol{\rho})=|\Psi_j(\boldsymbol{\rho})|^2$ is the equilibrium 2D density. The effective 2D interaction constants are}

\begin{align}
 g_{jj}&=\frac{\sqrt{8\pi}\hbar^2 a_{jj}}{m_j\ell_{z,j}},\qquad j=A,B,
 \label{eq:II_g_intra}\\
 g_{AB}=g_{BA}&=
 \frac{2\sqrt{\pi}\hbar^2 a_{AB}}
 {\mu_{AB}\sqrt{\ell_{z,A}^2+\ell_{z,B}^2}}.
 \label{eq:II_g_inter}
\end{align}
Here $a_{jj}$ and $a_{AB}$ are the intra- and interspecies $s$-wave scattering lengths, $\ell_{z,j}=\sqrt{\hbar/(m_j\omega_z)}$ is the axial harmonic-oscillator length, and $\mu_{AB}=m_Am_B/(m_A+m_B)$ is the reduced mass of an $A$--$B$ pair.

\textcolor{black}{At equilibrium, component $B$ occupies a central disk, while component $A$ forms the surrounding annulus.}
 We use $N_A=1.0\times10^4$, $N_B=1.5\times10^4$, $a_{AA}=99a_0$, and $a_{BB}=75a_0$, where $a_0$ is the Bohr radius.
The reference interspecies scattering length for the tension response and the eightfold-mode dynamics is $a_{AB}=426a_0$. \textcolor{black}{The corresponding segregation parameter is $K=g_{AB}/\sqrt{g_{AA}g_{BB}}=4.94397$.}
 For a uniform repulsive mixture, $K>1$ is the immiscibility condition; the trapped density distribution also depends on the atom numbers and confinement~\cite{Timmermans1998,Ao1998,Lee2016}. 

\textcolor{black}{For these parameters, Fig.~\ref{fig:equilibrium} shows a nearly circular internal interface between components $A$ and $B$, with an equilibrium radius $R_0=11.9631\,\mu\mathrm{m}$. The radial density profiles in Fig.~\ref{fig:equilibrium}(d) resolve the narrow region in which the components overlap.}

 This boundary separates two condensates, rather than a condensate from vacuum, so its motion involves both components.

\subsection{Interspecies-interaction drive}
\label{subsec:II_drive}

\begin{figure}[!t]
  \centering
  \includegraphics[width=0.99\columnwidth]{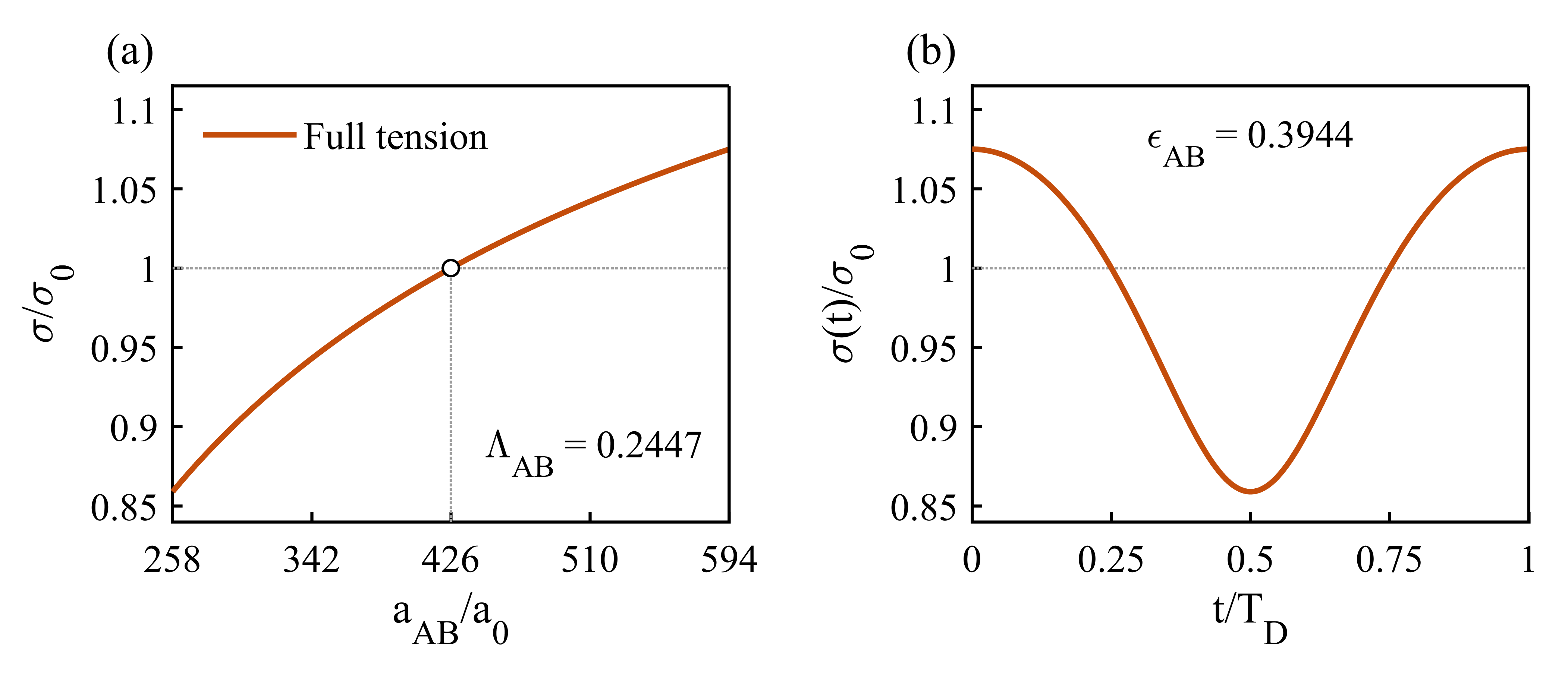}

\caption{Interaction dependence and periodic modulation of the interfacial tension.
(a) Normalized interfacial tension $\sigma/\sigma_0$ as a function of the interspecies scattering length $a_{AB}$. The open circle marks the reference point $a_{AB}=426a_0$, where the logarithmic tension sensitivity is $\Lambda_{AB}=0.2447$.
(b) Normalized interfacial tension $\sigma(t)/\sigma_0$ over one driving period $T_D$ for a relative modulation amplitude $\epsilon_{AB}=168/426\simeq0.3944$.
Both curves are calculated using the unexpanded tension function at fixed reference pressure and healing lengths.}
\label{fig:tension_response}
\end{figure}

We denote the modulation amplitude by $\Delta a_{AB}$ and its relative amplitude by $\epsilon_{AB}=\Delta a_{AB}/a_{AB}$. The driving angular frequency is $\omega_D=2\pi f_D$, where $f_D$ is the driving frequency and $T_D=1/f_D$ is the driving period. Only the interspecies scattering length is modulated, according to
\begin{equation}
 a_{AB}(t)=a_{AB}\left[1+\epsilon_{AB}\cos(\omega_Dt)\right].
 \label{eq:II_drive}
\end{equation}
Heteronuclear Feshbach resonances observed in this isotope mixture offer a possible basis for calibrated magnetic-field control of $a_{AB}$~\cite{Papp2006,Chin2010}. The intraspecies interaction constants and trapping potentials remain fixed. Equation~\eqref{eq:II_g_inter} implies $g_{AB}(t)=g_{AB}[1+\epsilon_{AB}\cos(\omega_Dt)]$ and $K(t)=K[1+\epsilon_{AB}\cos(\omega_Dt)]$. For the tension-response curve in Fig.~\ref{fig:tension_response}, $\Delta a_{AB}=168a_0$ and $\epsilon_{AB}=0.394366$, so $a_{AB}(t)$ spans $258a_0$ to $594a_0$ and $K(t)>1$ throughout the cycle. The spatially uniform modulation preserves rotational symmetry and imposes neither an azimuthal order $m$ nor a preferred orientation.

During the driven evolution, $n_j(\boldsymbol{\rho},t)=|\Psi_j(\boldsymbol{\rho},t)|^2$ denotes the time-dependent 2D density. The drive acts through the overlap of the two components. The interspecies interaction energy is
\begin{equation}
 E_{AB}(t)=g_{AB}(t)\int d^2\boldsymbol{\rho}\,
 n_A(\boldsymbol{\rho},t)n_B(\boldsymbol{\rho},t).
 \label{eq:II_overlap_energy}
\end{equation}
At fixed instantaneous fields, a change $\delta g_{AB}$ changes the local interaction-energy density by $\delta g_{AB}n_An_B$. In the strongly phase-separated state, this spatial weight is concentrated near the internal interface~\cite{Ao1998,Barankov2002}. The interaction dependence of the overlap layer therefore provides a microscopic origin for a time-dependent interfacial tension. The corresponding real-time GP dynamics are given in Appendix~A, Eq.~\eqref{eq:app_time_GP}.

\subsection{Interfacial-tension response and modulation sensitivity}
\label{subsec:II_tension}

Interfacial tension converts the interaction modulation into a modulation of the capillary restoring force. In the effective 2D model, $\sigma$ is the excess grand potential per unit interface length and has units of $\mathrm{J/m}$~\cite{Barankov2002,Indekeu2015}. For a boundary with uniform tension and length $L$, its interfacial contribution is $E_\sigma=\sigma L$. An increase in interface length thus carries an energetic cost. We describe the local interface structure with the planar-interface approximation of Van Schaeybroeck~\cite{VanSchaeybroeck2008}, expressed in effective 2D quantities as
\begin{equation}
 \sigma(a_{AB})=4P\xi_B F(\chi,K),\qquad
 \chi=\frac{\xi_A}{\xi_B},
 \label{eq:II_sigma}
\end{equation}
with the adopted planar-tension approximation
\begin{equation}
 \begin{aligned}
 F(\chi,K)={}&\frac{\sqrt{2}}{3}(1+\chi)
 -0.514\frac{\sqrt{\chi}}{K^{1/4}}\\
 &-\sqrt{\chi}\left(\chi+\frac{1}{\chi}\right)
 \left(\frac{0.055}{K^{3/4}}+\frac{0.067}{K^{5/4}}\right).
 \end{aligned}
 \label{eq:II_F}
\end{equation}
Here $P$ is the common pressure of the planar reference phases, $\xi_j$ are their healing lengths, and $\chi$ is the healing-length ratio. The reference pressure is the arithmetic mean of the two bulk pressures obtained by local Thomas--Fermi continuation to $R_0$. The reference bulk densities on the two sides of the interface are $n_j=[\mu_j-V_j(R_0)]/g_{jj}$, where $\mu_j$ are the chemical potentials of the undriven GP state, and $P=\frac14\sum_j g_{jj}n_j^2$. The corresponding planar coexistence densities $n_{j,\mathrm p}=\sqrt{2P/g_{jj}}$ define $\xi_j=\hbar/\sqrt{2m_jg_{jj}n_{j,\mathrm p}}$. These reference quantities are held fixed when evaluating the dependence on $a_{AB}$.

The logarithmic sensitivity quantifies the relative tension change produced by a small relative change of the scattering length. At the reference interaction, define
\begin{equation}
 \begin{aligned}
 \Lambda_{AB}
 &=\left.\frac{\partial\ln\sigma}{\partial\ln a_{AB}}
 \right|_{P,\xi_A,\xi_B}\\
 &=\frac{K}{F(\chi,K)}
 \frac{\partial F(\chi,K)}{\partial K}.
 \end{aligned}
 \label{eq:II_sensitivity}
\end{equation}
The derivative is evaluated at the reference scattering length $a_{AB}$. The second equality follows from $K\propto a_{AB}$ at fixed intraspecies interaction constants and axial profiles. The reference-state evaluation yields $\sigma_0\equiv\sigma[a_{AB}]=2.2708\times10^{-24}\,\mathrm{J/m}$ and $\Lambda_{AB}=0.24469$. A small $1\%$ increase in the scattering length therefore increases the tension by approximately $0.245\%$.
The open circle in Fig.~\ref{fig:tension_response}(a) marks the reference state, where $\Lambda_{AB}$ specifies the local logarithmic slope of the tension curve.

The relative scattering-length and tension modulations are connected by
\begin{equation}
 \begin{aligned}
 \frac{\sigma(t)}{\sigma_0}
 &=1+\eta_{AB}\cos(\omega_Dt)+O(\epsilon_{AB}^2),\\
 \eta_{AB}&=\Lambda_{AB}\epsilon_{AB}.
 \end{aligned}
 \label{eq:II_linear_tension}
\end{equation}
The dimensionless coefficient $\eta_{AB}$ is the first-order relative amplitude of the tension modulation. For $\epsilon_{AB}=168/426$, $\eta_{AB}=0.09650$: a $39.4\%$ relative scattering-length modulation corresponds to a $9.65\%$ relative tension modulation at first order. The sensitivity thus determines the effective capillary driving strength associated with the imposed interaction modulation.

{\color{black}
Magnetic Feshbach tuning provides a connection between the calculated interfacial-tension sensitivity and an applied magnetic field. We consider a bias field $B_{\mathrm b}$ satisfying $a_{AB}(B_{\mathrm b})=a_{AB}$. With the reference pressure, healing lengths, and intraspecies interaction constants held fixed, the relative tension response to the magnetic field is characterized by

\begin{equation}
S_B=\frac{\Lambda_{AB}}{a_{AB}}
\left.\frac{da_{AB}}{dB}\right|_{B_{\mathrm b}}.
\label{eq:II_magnetic_sensitivity}
\end{equation}

For a small sinusoidal field modulation $B(t)=B_{\mathrm b}+\delta B\cos(\omega_Dt)$, the first-order tension response is

\begin{equation}
\frac{\sigma(t)}{\sigma_0}
\simeq 1+S_B\delta B\cos(\omega_Dt).
\label{eq:II_magnetic_tension}
\end{equation}

Thus, $|S_B|\delta B$ determines the first-order relative modulation amplitude of the capillary restoring force, while the sign of $S_B$ determines its phase relative to the applied field. For larger field excursions, the tension waveform follows from $\sigma[a_{AB}(B(t))]$, retaining the nonlinear dependence of the scattering length on the magnetic field and of the tension on the scattering length.
}

Within the same fixed-reference tension model, the finite-amplitude response in Fig.~\ref{fig:tension_response}(b) follows by evaluating the same tension function along the prescribed interaction cycle,
\begin{equation}
 \frac{\sigma(t)}{\sigma_0}
 =\frac{F[\chi,K(t)]}{F(\chi,K)}.
 \label{eq:II_nonlinear_tension}
\end{equation}
The tension curve is increasing and concave over the displayed interaction range. Equal positive and negative changes of $a_{AB}$ therefore produce unequal changes of $\sigma$: the tension decrease is larger than the increase. The resulting periodic tension has a shifted cycle average and higher temporal harmonics even though the scattering length is sinusoidal. Its positive value throughout the cycle means that the drive modulates the strength of the capillary restoring force rather than reversing its sign. At fixed interface radius and mode inertia, a small $m$-fold deformation inherits this periodic restoring coefficient. The nonlinear interaction response therefore enters the parametric excitation problem already at linear order in the interface displacement.

\section{Interfacial-mode dispersion and Floquet instability}
\label{sec:capillary_floquet}

\subsection{Interface-mode equation and dispersion relation}
\label{subsec:III_modes}

\textcolor{black}{Small deformations of the nearly circular internal interface can be resolved into azimuthal modes.}
 For a standing mode with a fixed angular orientation, we write
\begin{equation}
 r(\theta,t)=R_0+\zeta_m(t)\cos(m\theta+\phi_m),
 \label{eq:III_deformation}
\end{equation}

where $\zeta_m(t)$ is the signed amplitude of the $m$-fold interface deformation, $m$ is the azimuthal mode number, and $\phi_m$ is a fixed angular phase.
 The angular dependence repeats under $\theta\to\theta+2\pi/m$ and gives the interface its $m$-fold rotational symmetry. The following calculation applies to each standing-wave component of the linear deformation.

Both condensates contribute to the inertia of the interface~\cite{Sasaki2009,Takeuchi2010}. 
We adopt an incompressible, irrotational sharp-interface model with fixed radius $R_0$ and reference bulk densities $n_A$ and $n_B$ on the two sides of the interface.
 Introducing velocity potentials through $\boldsymbol v_j=\nabla_\perp\Phi_j$, the continuity equation gives $\nabla_\perp^2\Phi_j=0$. For component $B$, regularity at $r=0$ selects the radial dependence $r^m$; for component $A$, the solution decaying away from the interface has radial dependence $r^{-m}$. The common normal-velocity condition at $r=R_0$ then gives
\begin{equation}
 \begin{aligned}
 \Phi_B&=\frac{R_0\dot\zeta_m}{m}
 \left(\frac{r}{R_0}\right)^m\cos(m\theta+\phi_m),\\
 \Phi_A&=-\frac{R_0\dot\zeta_m}{m}
 \left(\frac{R_0}{r}\right)^m\cos(m\theta+\phi_m).
 \end{aligned}
 \label{eq:III_potentials}
\end{equation}
The linear Bernoulli relation, $\delta P_j=-m_jn_j\partial_t\Phi_j$, yields
\begin{equation}
 \delta P_B-\delta P_A
 =-\frac{R_0}{m}(m_An_A+m_Bn_B)
 \ddot\zeta_m\cos(m\theta+\phi_m).
 \label{eq:III_pressure}
\end{equation}
Here $\delta P_j$ is the pressure perturbation of component $j$, and $m_jn_j$ is its 2D mass density. The sum of the two 2D mass densities reflects the flow induced on both sides of the common interface.

The interfacial tension supplies the restoring stress through the curvature variation. To linear order in the deformation,
\begin{equation}
 \delta\kappa_m=\frac{m^2-1}{R_0^2}
 \zeta_m\cos(m\theta+\phi_m),
 \label{eq:III_curvature}
\end{equation}
where $\delta\kappa_m$ is the curvature perturbation. The capillary contribution to the linearized normal-stress balance gives
\begin{equation}
 \frac{R_0}{m}(m_An_A+m_Bn_B)\ddot\zeta_m
 +\frac{\sigma(t)}{R_0^2}(m^2-1)\zeta_m=0.
 \label{eq:III_balance}
\end{equation}
Substituting the first-order tension response from Sec.~II gives the Mathieu equation
\begin{equation}
 \ddot\zeta_m+\omega_m^2
 [1+\eta_{AB}\cos(\omega_Dt)]\zeta_m=0,
 \label{eq:III_Mathieu}
\end{equation}
where $\eta_{AB}=\Lambda_{AB}\epsilon_{AB}$ is the relative tension-modulation amplitude at first order, and
\begin{equation}
 \omega_m^2=\frac{\sigma_0}{R_0^3}
 \frac{m(m^2-1)}{m_An_A+m_Bn_B}
 \label{eq:III_dispersion}
\end{equation}
is the capillary dispersion relation of the azimuthal interface modes in this sharp-interface model. Setting the modulation amplitude to zero reduces Eq.~\eqref{eq:III_Mathieu} to a harmonic oscillator with natural angular frequency $\omega_m$. The curvature factor $m^2-1$ and the mode-dependent fluid inertia produce the increase of frequency with azimuthal order. The primary parametric resonance therefore moves to higher driving frequencies for higher-order modes, with $\omega_D\simeq2\omega_m$~\cite{Staliunas2002,Maity2020,Kwon2021}.

At finite interaction-modulation amplitude, retaining the unexpanded tension dependence gives the periodic interface equation from Eq.~\eqref{eq:III_balance} as
\begin{equation}
 \ddot\zeta_m+
 \omega_m^2\frac{\sigma[a_{AB}(t)]}{\sigma_0}\zeta_m=0.
 \label{eq:III_periodic_mode}
\end{equation}
This is a Hill equation with the tension waveform specified in Sec.~II~\cite{Benjamin1954,Staliunas2002}. It retains the cycle-average shift and higher temporal harmonics generated by the nonlinear interaction dependence of the tension.

\subsection{Natural frequency from free GP oscillations}
\label{subsec:III_free}

Free GP oscillations provide an independent check of the natural frequency in Eq.~\eqref{eq:III_dispersion}. For $m=8$, this dispersion relation gives
\begin{equation}
 f_8=\frac{\omega_8}{2\pi}=47.87566\,\mathrm{Hz},
 \label{eq:III_f8}
\end{equation}
using the static interface parameters of Sec.~II. The GP calculation starts from a weak eightfold deformation and evolves at constant $a_{AB}=426a_0$, without periodic modulation. We resolve the interface motion through its complex angular Fourier coefficient,
\begin{equation}
 q_m(t)=\frac{1}{2\pi}\int_0^{2\pi}
 [r(\theta,t)-R(t)]e^{-im\theta}\,d\theta,
 \label{eq:III_q}
\end{equation}
where $R(t)$ is the instantaneous angular mean of the interface radius. The eightfold deformation coordinate is
\begin{equation}
 X_8(t)=2\operatorname{Re}\!\left[q_8(t)e^{-i\phi_0}\right],
 \qquad \phi_0=\arg q_8(0).
 \label{eq:III_X8}
\end{equation}
The fixed initial phase retains the alternating displacement of the interface, rather than its absolute magnitude.

\begin{figure}[!t]
  \centering
  \includegraphics[width=0.99\columnwidth]{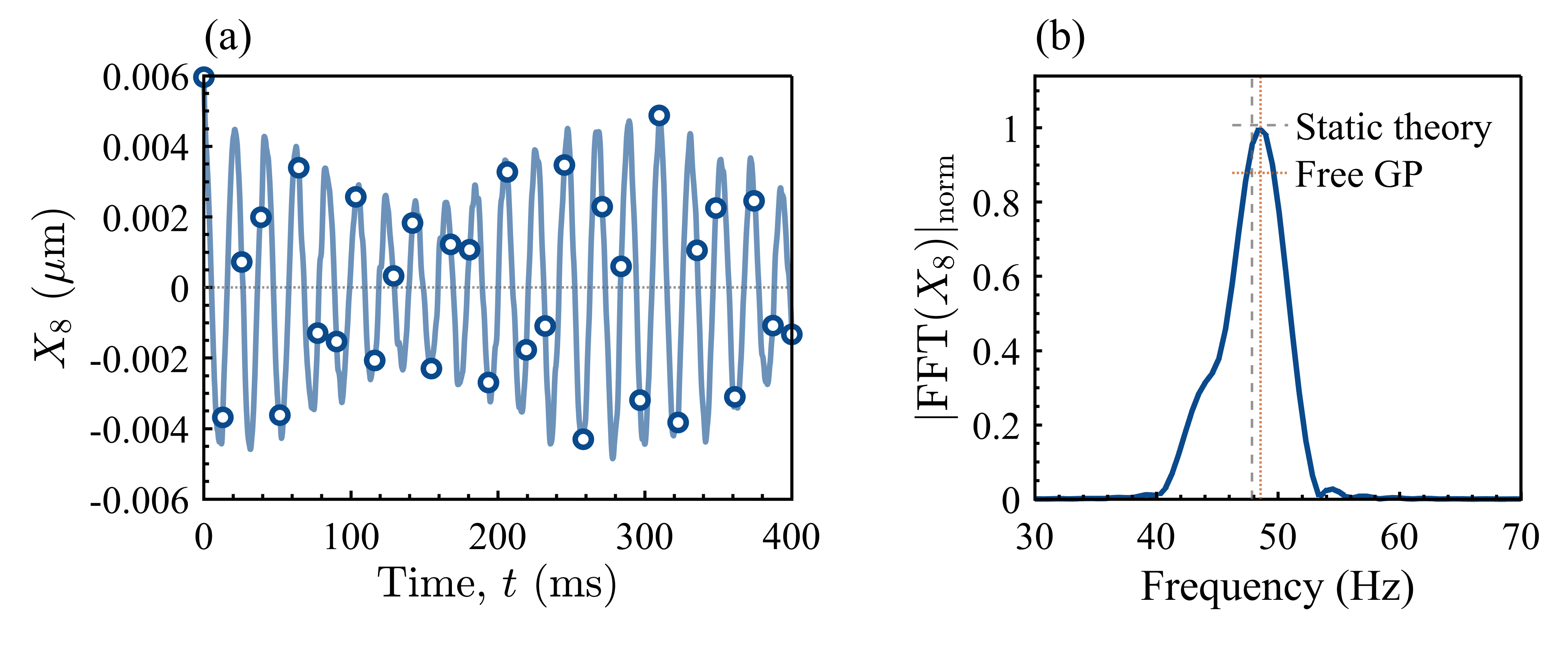}
  
\caption{Free oscillation of the eightfold interface mode at constant $a_{AB}=426a_0$.
(a) Phase-sensitive eightfold interface displacement $X_8(t)$ obtained from the GP evolution.
(b) Fourier amplitude spectrum normalized by its maximum. The gray dashed line marks the capillary prediction $f_8=47.8757\,\mathrm{Hz}$, and the orange dotted line marks the frequency obtained from the free GP oscillation, $48.56\,\mathrm{Hz}$.}

  \label{fig:free_ringdown}
\end{figure}

Fig.~\ref{fig:free_ringdown} shows the free oscillation and its temporal amplitude spectrum. The initial deformation satisfies $2|q_8(0)|/R_0\simeq4.98\times10^{-4}$. The principal frequency extracted from the GP trajectory is approximately $48.56\,\mathrm{Hz}$, close to the value given by Eq.~\eqref{eq:III_dispersion}. The comparison supports the natural-frequency scale of the eightfold mode and places its primary parametric resonance near twice this frequency.

\subsection{Floquet instability and parametric amplification}
\label{subsec:III_floquet}

Floquet analysis determines the instability of the periodic interface equation. Classical two-fluid interfaces have likewise been treated with Floquet methods, including viscous effects absent from the present capillary model~\cite{KumarTuckerman1994,Hill2002}. Let $M_m$ propagate the state $(\zeta_m,\dot\zeta_m)$ over one driving period $T_D$, with Floquet multipliers $\lambda_{m,+}$ and $\lambda_{m,-}$. We denote the Floquet amplitude growth rate by
\begin{equation}
 \textcolor{black}{\Gamma_m}=\frac{1}{T_D}
 \max_{\nu=\pm}\ln|\lambda_{m,\nu}|.
 \label{eq:III_Gamma}
\end{equation}
The absence of a first-derivative term in Eq.~\eqref{eq:III_periodic_mode} gives $\det M_m=1$. Inside the primary subharmonic tongue, the multipliers are negative real numbers and are reciprocal. For the individual Floquet mode associated with the multiplier of modulus greater than one,
\begin{equation}
 \zeta_m(t+T_D)=-e^{\textcolor{black}{\Gamma_m} T_D}\zeta_m(t).
 \label{eq:III_flip}
\end{equation}
After removal of the exponential envelope, this mode is antiperiodic over $T_D$ and periodic over $2T_D$. Its angular maxima and minima exchange positions after one driving period, while their magnitudes increase. For the angular dependence in Eq.~\eqref{eq:III_deformation}, this exchange is equivalent to an angular shift of $\pi/m$~\cite{Maity2020}. The marginal subharmonic solution has multiplier $-1$, and the corresponding tongue boundaries satisfy $\operatorname{tr}M_m=-2$.

\begin{figure}[!t]
  \centering
  
\includegraphics[width=0.99\columnwidth]
{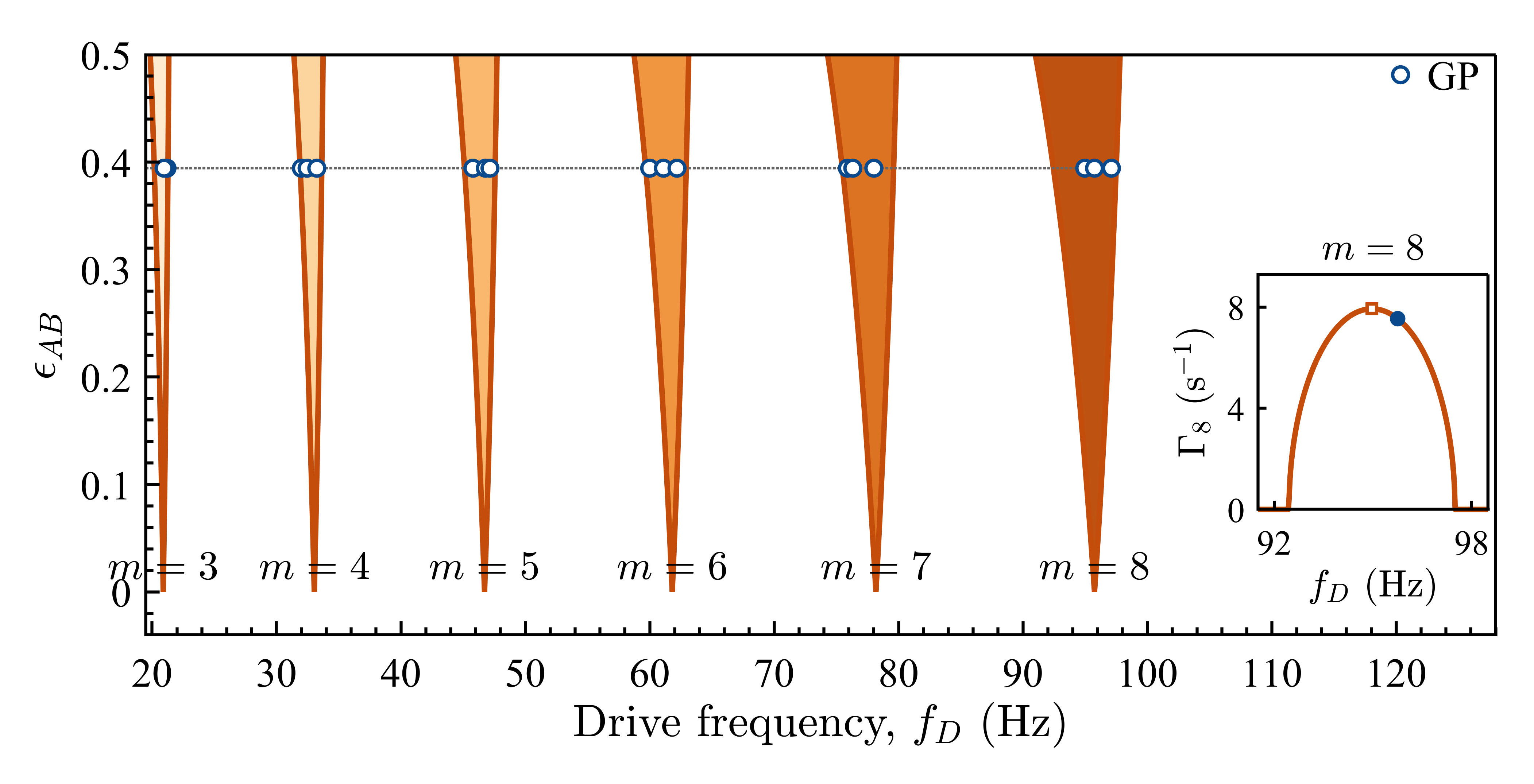}
  
  \caption{Primary Floquet instability tongues for interfacial modes $m=3$--$8$, plotted against driving frequency $f_D$ and relative modulation amplitude $\epsilon_{AB}$ of $a_{AB}$. The orange regions show the predictions of the capillary model. Open circles mark GP driving parameters at $\epsilon_{AB}=168/426$. 
The inset shows the theoretical Floquet amplitude growth rate $\textcolor{black}{\Gamma_8}$ as a function of $f_D$ at $\epsilon_{AB}=168/426\simeq0.3944$. The open square marks its maximum, and the filled circle marks its value at $f_D=95.7513\,\mathrm{Hz}$.
}
  \label{fig:floquet}
\end{figure}

Fig.~\ref{fig:floquet} shows the primary instability tongues for \mbox{$m=3$--$8$}. Their zero-amplitude limits occur at $f_D=2f_m$, where $f_m=\omega_m/(2\pi)$, and move to higher frequencies with increasing $m$. For weak modulation, the tongue widths are controlled by $\eta_{AB}=\Lambda_{AB}\epsilon_{AB}$. The dispersion relation sets the frequency ordering, whereas the tension sensitivity converts the imposed scattering-length modulation into the effective capillary drive. At finite modulation amplitude, the cycle-average shift and higher temporal harmonics of $\sigma[a_{AB}(t)]$ also modify the tongue positions and shapes. The open circles mark the driving parameters used in the GP simulations.

The inset resolves the frequency dependence of the eightfold-mode Floquet growth rate. At $\epsilon_{AB}=168/426$, the tension function without a first-order expansion gives the instability interval $92.4455<f_D<97.4947\,\mathrm{Hz}$. The maximum growth rate is $7.9308\,\mathrm{s^{-1}}$ at $f_D=94.9701\,\mathrm{Hz}$. At $f_D=95.7513\,\mathrm{Hz}$, the same periodic interface equation gives $\textcolor{black}{\Gamma_8}=7.5415\,\mathrm{s^{-1}}$. These quantities are temporal amplitude growth rates; the corresponding amplification factor over one driving cycle is $\exp(\textcolor{black}{\Gamma_m}T_D)$.

\begin{figure*}[!t]
  \centering
  \includegraphics[width=0.94\textwidth]{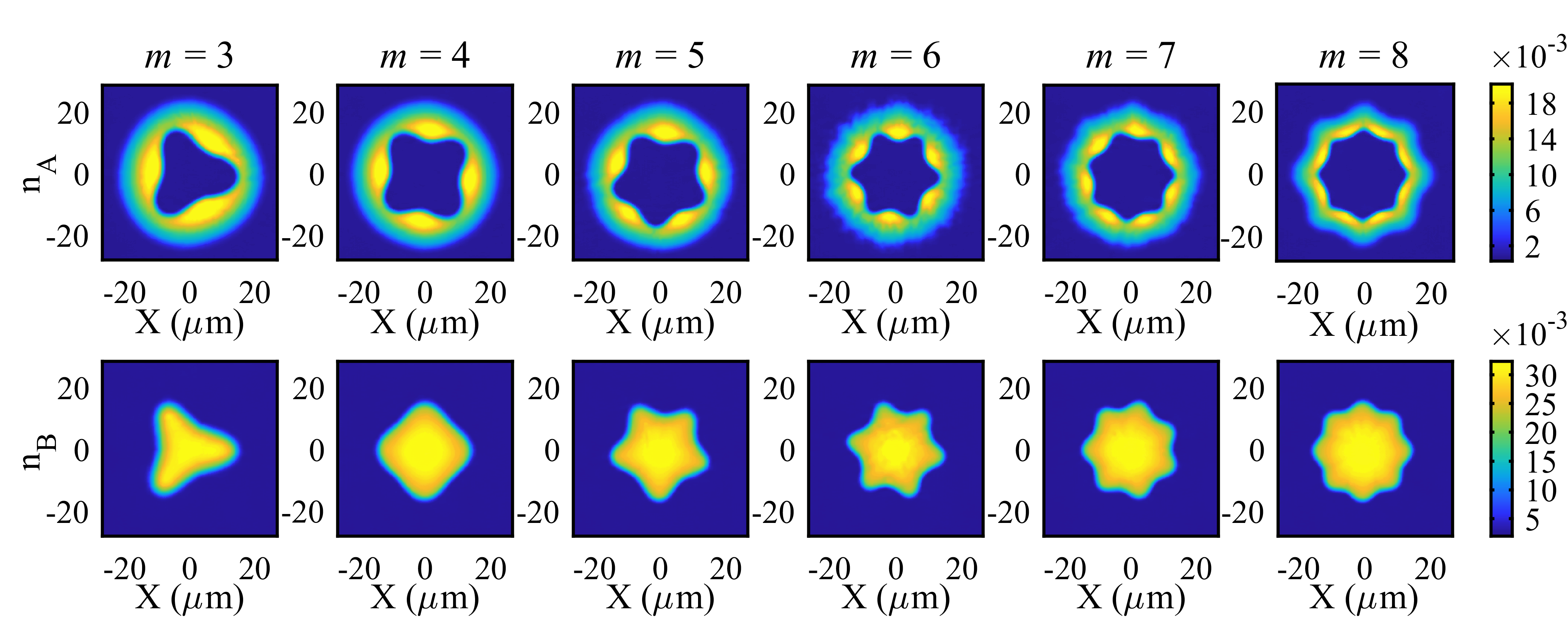}
  \caption{Star-shaped deformations of the internal $A$--$B$ interface with azimuthal orders $m=3$--$8$.
The driving frequencies from left to right are $f_D=21.19,\,32.46,\,47.15,\,62.20,\,78.59,\,97.16\,\mathrm{Hz}$.
 All snapshots use the reference scattering length $a_{AB}=426a_0$ and modulation amplitude $\Delta a_{AB}=168a_0$. The two components exhibit complementary deformations of their common internal interface.}
  \label{fig:stars}
\end{figure*}

The periodic modulation of the interfacial tension can transfer energy to the interface modes. Multiplying Eq.~\eqref{eq:III_periodic_mode} by $\dot\zeta_m$ gives

\begin{equation}
 \frac{d}{dt}\left[
 \frac{\dot\zeta_m^{\,2}}{2}
 +\frac{\omega_m^2\sigma(t)}{2\sigma_0}\zeta_m^2
 \right]
 =\frac{\omega_m^2}{2\sigma_0}\dot\sigma(t)\zeta_m^2.
 \label{eq:III_work}
\end{equation}

Integrating Eq.~\eqref{eq:III_work} over one driving period, the interface mode gains energy when

\begin{equation}
 \int_t^{t+T_D}\dot\sigma(t')\zeta_m^2(t')\,dt'>0.
 \label{eq:III_positive_work}
\end{equation}
Thus amplification depends on the relative phase of the tension modulation and the squared deformation. For a nearly sinusoidal mode, the squared deformation contains a component at $2\omega_m$, allowing coherent energy transfer near the primary parametric resonance. The positive Floquet exponent describes the cumulative amplification associated with this phase-dependent work. Because the interfacial tension remains positive throughout the drive cycle, the instability arises from phase-dependent energy transfer under periodic modulation, rather than from a reversal of the capillary restoring force.

\section{Star-shaped patterns and subharmonic interface motion}
\label{sec:patterns_subharmonic}

\subsection{Angular structure of the interface patterns}
\label{subsec:IV_patterns}

\textcolor{black}{Periodic modulation of the interspecies scattering length produces star-shaped deformations of the internal interface between components $A$ and $B$.}
 Fig.~\ref{fig:stars} shows density distributions with threefold to eightfold angular structure, corresponding to \mbox{$m=3$--$8$}. The upper and lower rows show components $A$ and $B$, respectively. 
As the driving frequency $f_D$ increases, the azimuthal order of the displayed interface patterns increases from $m=3$ to $m=8$. The corresponding angular periodicity is $2\pi/m$.

The paired density distributions show complementary deformations at the shared boundary. An outward protrusion of component $B$ is accompanied by a corresponding recession of component $A$, whereas an inward indentation of $B$ allows $A$ to extend toward the center. The two density distributions therefore resolve the same angular displacement from opposite sides of the interface.

This frequency ordering is consistent with the dispersion relation in Eq.~\eqref{eq:III_dispersion}.
 Within the fixed-reference capillary model, higher-order corrugations have a larger curvature-dependent restoring coefficient and a higher natural frequency. Their primary parametric resonances consequently occur at higher driving frequencies, giving the sequence of angular orders displayed in Fig.~\ref{fig:stars}. The Floquet tongues in Fig.~\ref{fig:floquet} describe this ordering at the level of the linear interface modes, while the density distributions show the resulting spatial patterns.

\subsection{Subharmonic motion of the eightfold mode}
\label{subsec:IV_subharmonic}

The driven eightfold mode exhibits an oscillation at approximately half the modulation frequency. Its time-dependent GP trajectory uses $a_{AB}=426a_0$, $\Delta a_{AB}=168a_0$, and $f_D=95.7513\,\mathrm{Hz}$, starting from the same weak eightfold deformation used for the free oscillation in Sec.~III. We describe its motion by the deformation coordinate $X_8(t)$ defined in Eq.~\eqref{eq:III_X8}. The fixed angular phase in this coordinate retains the alternating outward and inward displacement of the eightfold component.

\begin{figure}[!t]
  \centering
  \includegraphics[width=0.99\columnwidth]{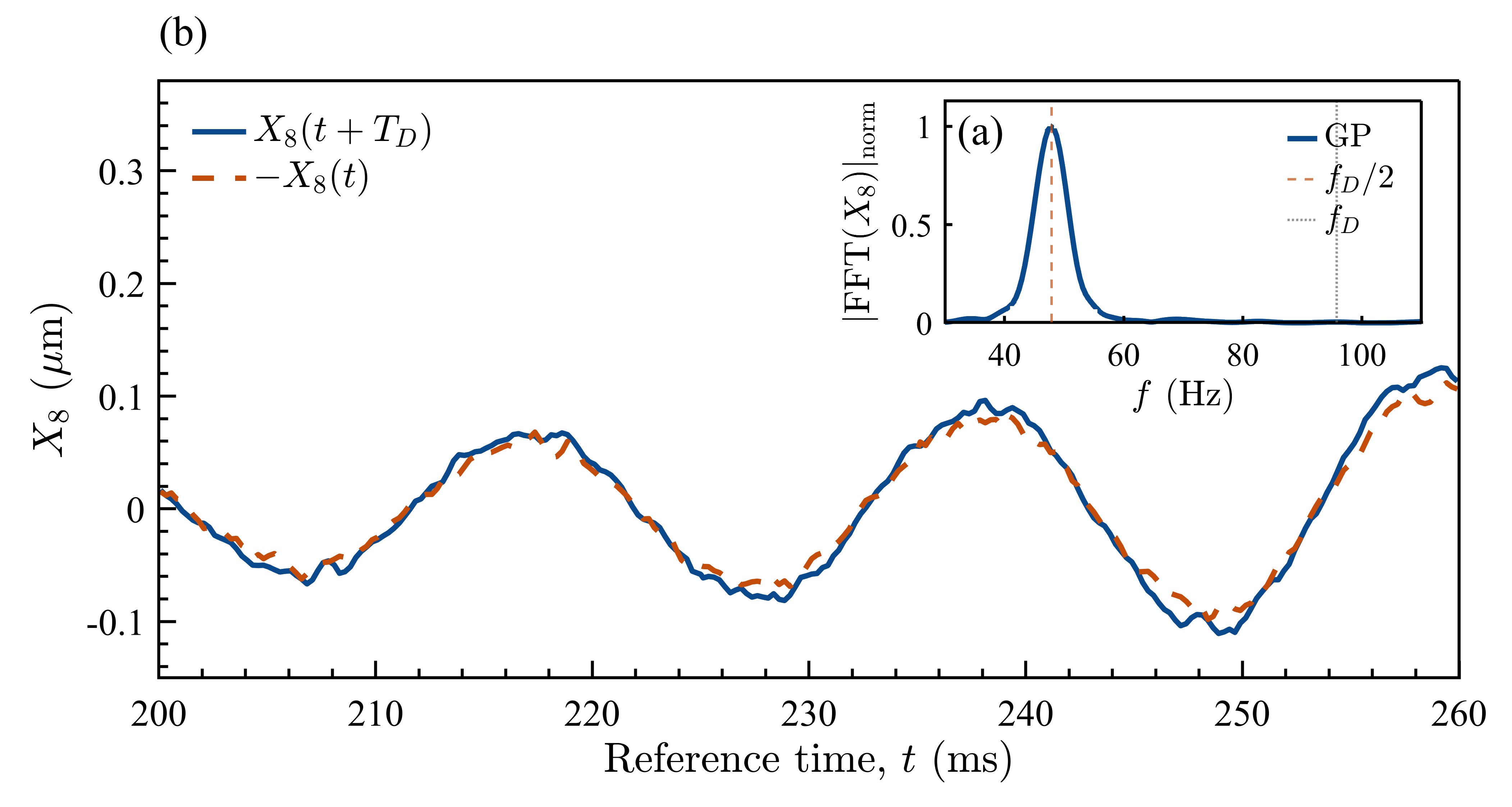}
  \caption{Subharmonic motion of the eightfold interface mode at $f_D=95.7513\,\mathrm{Hz}$. (a), inset: normalized Fourier amplitude spectrum of $X_8$, with reference lines at $f_D/2$ (orange dashed) and $f_D$ (gray dotted). (b), main panel: $X_8(t+T_D)$ (blue solid) and $-X_8(t)$ (orange dashed) for reference times $200$--$260\,\mathrm{ms}$.
The approximate overlap of the two curves indicates a phase shift of $\pi$ after one driving period.
}
  \label{fig:subharmonic}
\end{figure}

The temporal amplitude spectrum in the inset of Fig.~\ref{fig:subharmonic} has its principal peak near the half-frequency reference $f_D/2=47.87565\,\mathrm{Hz}$, with a much smaller contribution at $f_D$. The dominant oscillation thus has a period close to $2T_D=20.9\,\mathrm{ms}$. This half-frequency response is the temporal signature of the primary parametric resonance described by the interface-mode equation~\cite{Staliunas2002,Maity2020}.

The one-period phase relation is also resolved directly in the time domain. Fig.~\ref{fig:subharmonic} compares $X_8(t+T_D)$ with $-X_8(t)$ for reference times between $200$ and $260\,\mathrm{ms}$. Their extrema and zero crossings occur at nearly the same times, showing an approximately $\pi$ phase advance of the dominant oscillation over one driving period. The shifted signal generally has a larger magnitude, consistent with the concurrent growth of the deformation. The observed motion combines this alternating phase with an increasing amplitude, as expected for a growing subharmonic response.

For a standing interfacial wave, a half-cycle phase advance exchanges the angular positions of crests and troughs~\cite{Maity2020}. The angular dependence obeys
\begin{equation}
 -\cos(m\theta+\phi_m)
 =\cos\!\left[m\left(\theta+\frac{\pi}{m}\right)+\phi_m\right].
 \label{eq:IV_angular_exchange}
\end{equation}
For $m=8$, the two sets of crest directions are therefore separated by $\pi/8$. This is the alternating deformation of an eightfold standing wave. Its oscillatory phase approximately recurs after two driving periods, while its growing amplitude prevents the full deformation from repeating exactly after two driving periods.

\section{Interfacial amplification and azimuthal-mode dominance}
\label{sec:amplification_modes}

\subsection{Growth of the eightfold interface deformation}
\label{subsec:V_amplification}

The eightfold interfacial deformation grows under near-resonant modulation of the interspecies interaction. Its geometric amplitude is defined from the angular Fourier coefficient in Eq.~\eqref{eq:III_q} as
\begin{equation}
 A_m(t)=2|q_m(t)|,
 \label{eq:V_amplitude}
\end{equation}
where $A_m$ measures the magnitude of the $m$th angular deformation and is distinct from the imposed modulation amplitude $\Delta a_{AB}$. We denote its average over one driving period $T_D$ by $\overline A_m(t)$. The period-averaged amplitude characterizes the slowly evolving deformation envelope, while the phase-sensitive coordinate $X_8(t)$ in Sec.~IV describes the alternating displacement of the interface.

\begin{figure}[!t]
  \centering
  \includegraphics[width=0.99\columnwidth]{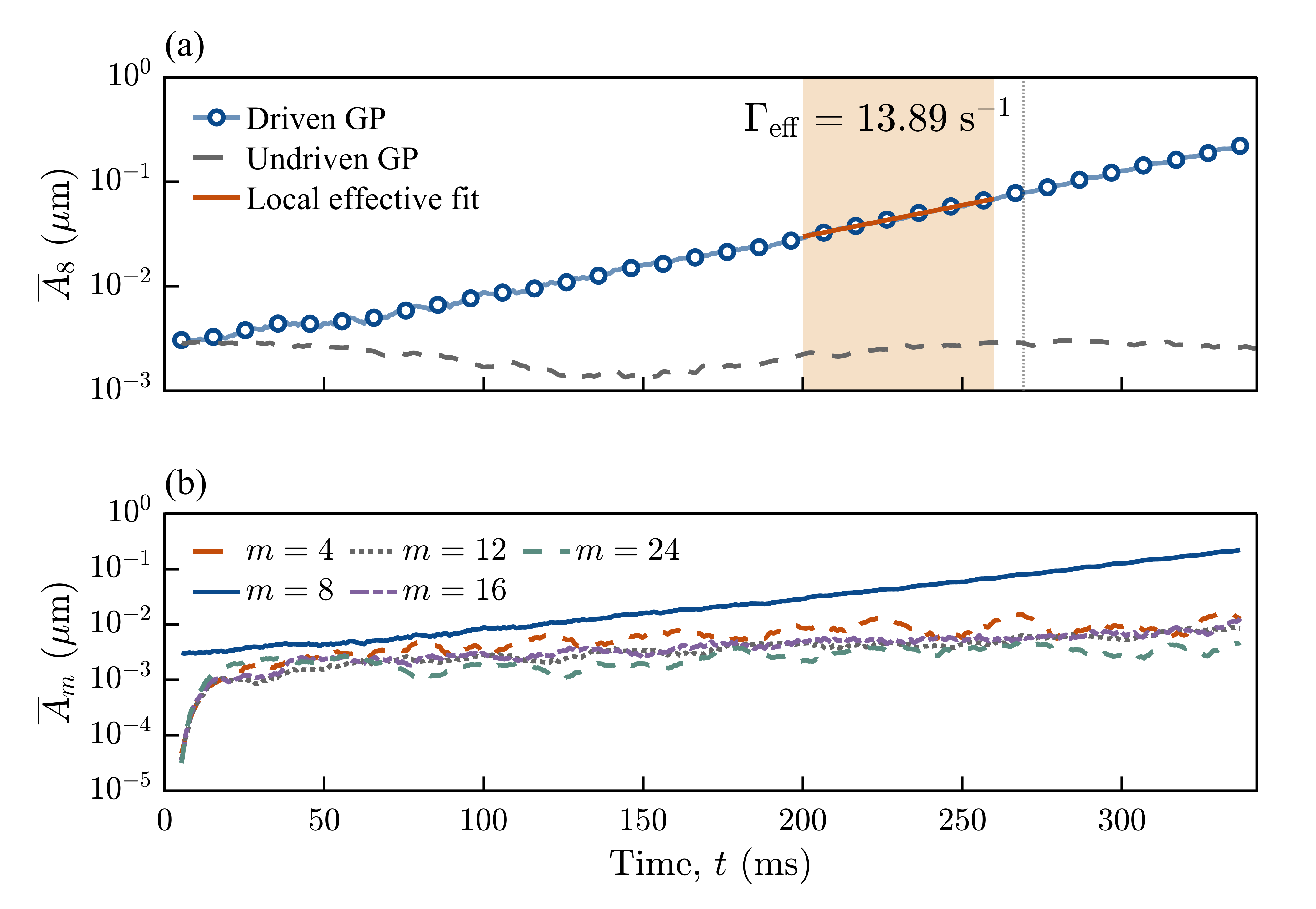}
  \caption{Eightfold-mode amplification and azimuthal-mode amplitudes at $f_D=95.7513\,\mathrm{Hz}$. (a) One-period-averaged eightfold geometric amplitude. The blue line and circles show driven GP evolution; the gray dashed line shows undriven evolution from the same initial deformation. The orange line is a local exponential fit over the shaded $200$--$260\,\mathrm{ms}$ interval, with $\Gamma_{\rm eff}\simeq13.9\,\mathrm{s^{-1}}$. The vertical dotted line marks the first raw $A_8/R=1\%$ crossing. (b) Period-averaged amplitudes for $m=8$ (blue solid), $m=4$ (orange dashed), $m=12$ (gray dotted), $m=16$ (purple dash-dotted), and $m=24$ (green dashed). Both panels use logarithmic amplitude axes.}
  \label{fig:amplification}
\end{figure}

The driven evolution exhibits a sustained increase in $\overline A_8$, whereas the undriven condensate, prepared with the same weak eightfold deformation, remains in a small-amplitude oscillatory state [Fig.~\ref{fig:amplification}(a)]. The separation between these trajectories identifies the increasing corrugation as a response to the periodic interaction drive. Its angular structure is inherited from the initially prepared mode, while its amplification is consistent with the instability of the eightfold capillary mode at the chosen driving frequency.

Over the small-deformation interval $200$--$260\,\mathrm{ms}$, the period-averaged amplitude follows an approximately exponential increase,

\begin{equation}
\overline A_8(t)\simeq
\overline A_8(\textcolor{black}{t^{*}})
\exp\!\left[
\Gamma_{\mathrm{eff}}
(t-\textcolor{black}{t^{*}})
\right],
\label{eq:V_growth}
\end{equation}

with a local effective amplitude growth rate $\Gamma_{\mathrm{eff}}\simeq13.9\,\mathrm{s^{-1}}$. 
\textcolor{black}{Here $t^{*}$ denotes a reference time within the fitting interval.}

\textcolor{black}{The corresponding amplitude-doubling time is approximately $50\,\mathrm{ms}$, equivalent to about five driving periods. The period-averaged amplitude increases by a factor of approximately $2.3$ over the $60\,\mathrm{ms}$ interval.}

\subsection{Dominance of the eightfold mode}
\label{subsec:V_modes}

The growing deformation remains dominated by the eightfold angular component. Fig.~\ref{fig:amplification}(b) compares $\overline A_m$ for $m=4,8,12,16,$ and $24$. The eighth-order amplitude increases markedly, while the other displayed angular components remain smaller. In the same $200$--$260\,\mathrm{ms}$ interval, $\overline A_8$ exceeds the largest competing component in the analyzed angular spectrum by more than a factor of three. The angular structure of the evolving interface is therefore governed predominantly by the $m=8$ mode, rather than by a comparable superposition of several modes.

The imposed interaction modulation is rotationally symmetric, but the interface supports discrete modes with different natural frequencies. At $f_D=95.7513\,\mathrm{Hz}$, the eightfold mode lies within its predicted primary instability tongue [Fig.~\ref{fig:floquet}]. The evolution shown in Fig.~\ref{fig:amplification} demonstrates preferential amplification of the initially prepared eightfold deformation under this drive. Together with the spatial patterns in Fig.~\ref{fig:stars} and the subharmonic motion in Fig.~\ref{fig:subharmonic}, the mode-resolved growth identifies the development of an eightfold Faraday interface oscillation in the GP dynamics.

\section{Conclusion and outlook}
\label{sec:conclusion}

\textcolor{black}{Periodic modulation of the interspecies scattering length excites azimuthal deformations of the internal interface separating the central $B$ disk from the surrounding $A$ annulus in a phase-separated quasi-2D binary BEC.}
The explicitly time-dependent interaction energy is weighted by the density overlap $n_A n_B$, which is concentrated in the interfacial region. Within the quasi-2D capillary model, the interaction dependence of the interfacial tension maps the imposed scattering-length modulation onto a periodic restoring force. The logarithmic tension sensitivity, $\Lambda_{AB}=0.24469$, determines this mapping to first order. At finite modulation amplitude, the nonlinear tension response also shifts the cycle-averaged restoring coefficient and introduces higher temporal harmonics.

The resulting interface-mode equation yields a discrete capillary dispersion relation and primary Floquet instability tongues. For \mbox{$m=3$--$8$}, the resonance regions move to higher driving frequencies with increasing azimuthal order, consistent with the frequency-ordered sequence of star-shaped interfaces in the density distributions obtained from GP simulations. The complementary deformations of components $A$ and $B$ identify these patterns as deformations of their shared boundary. The interfacial tension sets the natural-frequency scale, while its sensitivity determines the first-order response of the instability bands to the imposed interaction modulation.

The eightfold mode resolves the time-dependent growth of an interfacial pattern. Free GP oscillations give a frequency near $48.56\,\mathrm{Hz}$, close to the $47.88\,\mathrm{Hz}$ value obtained from the capillary dispersion relation. Under near-resonant driving, the deformation oscillates at approximately half the driving frequency, with crests and troughs exchanging positions after one driving period. Its period-averaged amplitude increases approximately exponentially over a small-deformation interval, with a local effective growth rate of $13.9\,\mathrm{s^{-1}}$, corresponding to an amplitude-doubling time of about $50\,\mathrm{ms}$. The initially prepared eightfold component remains dominant during this stage.

The relation between interspecies-interaction modulation and interfacial tension offers a route to frequency-selective excitation of interface modes and to probing the interaction dependence of the capillary restoring force. \textcolor{black}{Moreover, the observed subharmonic motion of the eightfold interface mode raises the question of whether subsequent nonlinear dynamics can sustain robust period-doubled interfacial oscillations and support discrete time-crystalline order~\cite{Else2016,Khemani2016,Smits2018}. At a fixed driving period, the relative modulation amplitude $\epsilon_{AB}$ could serve as a control parameter for exploring a possible transition from drive-periodic interfacial motion to a phase with subharmonic temporal order~\cite{Yao2020,Zaletel2023}.}

\begin{acknowledgments}
This work is supported by the Joint Fund for Quantum Major Research Plan of the National Natural Science Foundation of China (Grant No. U25D8014), the Key Special Project of the National Key Research and Development Program of China (Grant No. 2025YFF0515201), the Natural Science Foundation of Shanghai (Grant No. 23ZR1418700).
\end{acknowledgments}
\appendix
\section{Quasi-2D reduction}
\label{app:two_dimensional}

The effective 2D interaction constants used in Sec.~II follow from the three-dimensional (3D) contact interaction after integrating over the tightly confined axial direction. Let $\widetilde\Psi_j(\boldsymbol\rho,z,t)$ denote the 3D condensate order parameter of component $j=A,B$. We factorize it as
\begin{equation}
 \widetilde\Psi_j(\boldsymbol\rho,z,t)
 =\Psi_j(\boldsymbol\rho,t)\varphi_j(z),
 \qquad \int_{-\infty}^{\infty}|\varphi_j(z)|^2\,dz=1,
 \label{eq:app_factorisation}
\end{equation}
where $\Psi_j$ is the 2D order parameter normalized to the atom number $N_j$. For harmonic axial confinement with angular frequency $\omega_z$, the species-dependent ground states are
\begin{equation}
 \varphi_j(z)=\frac{1}{\pi^{1/4}\ell_{z,j}^{1/2}}
 \exp\!\left(-\frac{z^2}{2\ell_{z,j}^2}\right),
 \qquad \ell_{z,j}=\sqrt{\frac{\hbar}{m_j\omega_z}}.
 \label{eq:app_axial}
\end{equation}
The 3D contact interaction constant is
\begin{equation}
 g_{jk}^{\mathrm{3D}}=\frac{2\pi\hbar^2a_{jk}}{\mu_{jk}},
 \qquad \mu_{jk}=\frac{m_jm_k}{m_j+m_k}.
 \label{eq:app_3D_coupling}
\end{equation}
Here $m_j$ denotes the atomic mass, $a_{jk}$ the $s$-wave scattering length, and $\mu_{jk}$ the reduced mass. Inserting Eq.~\eqref{eq:app_factorisation} into the 3D GP equations and integrating over $z$ gives the effective 2D interaction constant
\begin{align}
 g_{jk}&=g_{jk}^{\mathrm{3D}}
 \int_{-\infty}^{\infty}|\varphi_j(z)|^2|\varphi_k(z)|^2\,dz\notag\\
 &=\frac{2\sqrt{\pi}\hbar^2a_{jk}}
 {\mu_{jk}\sqrt{\ell_{z,j}^2+\ell_{z,k}^2}}.
 \label{eq:app_2D_coupling}
\end{align}
For identical components, $\mu_{jj}=m_j/2$, hence
\begin{equation}
 g_{jj}=\frac{\sqrt{8\pi}\hbar^2a_{jj}}{m_j\ell_{z,j}}.
 \label{eq:app_2D_self}
\end{equation}
The axial zero-point energy is spatially uniform and may be removed by a phase transformation. For real-time evolution, the coupled 2D GP equations are
\begin{equation}
 \begin{aligned}
 i\hbar\frac{\partial\Psi_A}{\partial t}
 &=\biggl[-\frac{\hbar^2}{2m_A}\nabla_\perp^2
 +V_A(\boldsymbol\rho)\\
 &\quad+g_{AA}|\Psi_A|^2+g_{AB}(t)|\Psi_B|^2\biggr]\Psi_A,\\[2pt]
 i\hbar\frac{\partial\Psi_B}{\partial t}
 &=\biggl[-\frac{\hbar^2}{2m_B}\nabla_\perp^2
 +V_B(\boldsymbol\rho)\\
 &\quad+g_{BB}|\Psi_B|^2+g_{AB}(t)|\Psi_A|^2\biggr]\Psi_B,
 \end{aligned}
 \label{eq:app_time_GP}
\end{equation}
where $\Psi_j=\Psi_j(\boldsymbol\rho,t)$ satisfies $\int d^2\boldsymbol\rho\,|\Psi_j|^2=N_j$.
At fixed interaction constants, the stationary ansatz
\textcolor{black}{$\Psi_j(\boldsymbol\rho,t)=
\Psi_j(\boldsymbol\rho)e^{-i\mu_jt/\hbar}$}
reduces this equation to Eq.~\eqref{eq:II_GP}.

\textcolor{black}{The 2D density $n_j(\boldsymbol\rho,t)=|\Psi_j(\boldsymbol\rho,t)|^2$ is obtained by integrating the three-dimensional number density along the axial direction.} The interfacial tension in this effective 2D theory is an energy per unit interface length.

For the dimensionless propagation, we use $a_x=\sqrt{\hbar/(m_A\omega_\perp)}$ as the length scale and $\omega_\perp^{-1}$ as the time scale. With $\boldsymbol u=\boldsymbol\rho/a_x$, $\tau=\omega_\perp t$, and $\psi_j=a_x\Psi_j/\sqrt{N_j}$, the equations take the form
\begin{equation}
 i\partial_\tau\psi_j=\left[-\frac{m_A}{2m_j}\nabla_{\boldsymbol u}^2
 +\frac{m_j}{2m_A}u^2
 +\sum_{k=A,B}\beta_{jk}(\tau)|\psi_k|^2\right]\psi_j,
 \label{eq:app_dimensionless_GP}
\end{equation}
where $\int d^2\boldsymbol u\,|\psi_j|^2=1$ and $\beta_{jk}(\tau)=N_k g_{jk}(t)/(\hbar\omega_\perp a_x^2)$.
The ground state is obtained by imaginary-time propagation, followed by real-time GP evolution under the prescribed $a_{AB}(t)$.

\section{Reference interface, tension sensitivity, and Floquet construction}
\label{app:tension_floquet}

\subsection{Static inputs for the interfacial model}
\label{app:static_inputs}

The circular reference interface is located from the crossing of the two radial density profiles at the internal boundary, giving $R_0=11.9631\,\mu\mathrm m$. The reference densities $n_A$ and $n_B$ entering the mode inertia are the one-sided bulk densities extended to this radius. In the local Thomas--Fermi description of the undriven GP state, they are assigned by
\begin{equation}
 n_j=\frac{\mu_j-V_j(R_0)}{g_{jj}},\qquad j=A,B,
 \label{eq:app_bulk_densities}
\end{equation}
where $\mu_j$ is the chemical potential of component $j$. These values characterize the reference bulk phases, rather than the densities sampled directly inside the finite overlap layer.

The planar interfacial-tension expression requires a common coexistence pressure~\cite{VanSchaeybroeck2008,Indekeu2015}. We construct a local reference pressure from the two bulk-side pressures,
\begin{equation}
 P_j=\frac{g_{jj}n_j^2}{2},\qquad
 P=\frac{P_A+P_B}{2},
 \label{eq:app_pressure_reference}
\end{equation}
and use the associated planar coexistence densities and healing lengths,
\begin{equation}
 n_{j,\mathrm p}=\sqrt{\frac{2P}{g_{jj}}},\qquad
 \xi_j=\frac{\hbar}{\sqrt{2m_jg_{jj}n_{j,\mathrm p}}}.
 \label{eq:app_planar_density}
\end{equation}
This pressure construction separates the density input to the interfacial inertia from the equal-pressure reference used in the planar tension formula. All three reference quantities $P$, $\xi_A$, and $\xi_B$ are fixed when the interaction dependence of $\sigma$ is evaluated.

In the adopted tension expression $\sigma=4P\xi_BF(\chi,K)$, where $\chi=\xi_A/\xi_B$ and $K=g_{AB}/\sqrt{g_{AA}g_{BB}}$, the logarithmic sensitivity can be evaluated analytically. For the function $F(\chi,K)$ given in Sec.~II,
\begin{equation}
\begin{aligned}
 \frac{\partial F}{\partial K}={}&
 \frac{0.514\sqrt\chi}{4K^{5/4}} \\
 &+\sqrt\chi\left(\chi+\frac1\chi\right)
 \left[\frac{3(0.055)}{4K^{7/4}}+
       \frac{5(0.067)}{4K^{9/4}}\right].
\end{aligned}
 \label{eq:app_F_derivative}
\end{equation}
Thus $\Lambda_{AB}=K(\partial F/\partial K)/F$ at the reference scattering length. The static inputs give $K=4.94397$, $\chi=0.92484$, $\sigma_0=2.2708\times10^{-24}\,\mathrm{J/m}$, and $\Lambda_{AB}=0.24469$.

\subsection{Finite-amplitude tension modulation}
\label{app:nonlinear_tension}

At fixed reference pressure and healing lengths, the imposed sinusoidal drive produces the dimensionless tension coefficient
\begin{equation}
 h(t)=\frac{\sigma[a_{AB}(t)]}{\sigma_0}
 =\frac{F\!\left(\chi,K[1+\epsilon_{AB}\cos(\omega_Dt)]\right)}
 {F(\chi,K)}.
 \label{eq:app_h_full}
\end{equation}
The first-order coefficient is $h(t)\simeq1+\eta_{AB}\cos(\omega_Dt)$, with $\eta_{AB}=\Lambda_{AB}\epsilon_{AB}$. The origin of the mean shift and higher temporal harmonics can be seen directly by expanding about $a_{AB}$:
\begin{equation}
\begin{aligned}
 \sigma[a_{AB}(t)]={}&\sigma_0
 +\sigma'(a_{AB})\Delta a_{AB}\cos(\omega_Dt)\\
 &+\frac{\sigma''(a_{AB})(\Delta a_{AB})^2}{4}
 [1+\cos(2\omega_Dt)]+\cdots .
\end{aligned}
 \label{eq:app_tension_second_order}
\end{equation}
Here the primes denote derivatives with respect to the interspecies scattering length at fixed $P$ and $\xi_j$. The unexpanded function in Eq.~\eqref{eq:app_h_full}, rather than the truncated series in Eq.~\eqref{eq:app_tension_second_order}, is used for the finite-amplitude Floquet curves.

\ifdefined\IncludeAppendixTensionFigure
\begin{figure}[tb]
 \centering
 \includegraphics[width=\columnwidth]{figs/Appendix_tension_first_order_comparison.png}
 \caption{Interfacial-tension response from the adopted static tension function (solid) and its first-order expansion about $a_{AB}=426a_0$ (dashed). (a) Interaction dependence; the reference-point tangent defines $\Lambda_{AB}$. (b) Response during one driving period at $\epsilon_{AB}=168/426$. }
 \label{fig:app_tension}
\end{figure}
\fi

\subsection{Floquet calculation for the periodic interface mode}
\label{app:floquet_method}

The $m$th interfacial-mode equation is
\begin{equation}
 \ddot\zeta_m+\omega_m^2 h(t)\zeta_m=0,
 \qquad h(t+T_D)=h(t),
 \label{eq:app_Hill}
\end{equation}
where $\omega_m$ is obtained from the capillary-mode dispersion relation in Sec.~III. For the first-order tension response, the primary subharmonic Floquet expansion may be written as~\cite{Benjamin1954,Nicolin2007,Maity2020}
\begin{equation}
 \zeta_m(t)=e^{st}
 \sum_{p=-\infty}^{\infty}c_m^{(p)}
 e^{i(p+1/2)\omega_Dt}.
 \label{eq:app_Floquet_series}
\end{equation}
Insertion into Eq.~\eqref{eq:app_Hill} with $h(t)=1+\eta_{AB}\cos(\omega_Dt)$ gives the Mathieu--Floquet recurrence
\begin{equation}
\begin{aligned}
 &\left\{[s+i(p+\tfrac12)\omega_D]^2+\omega_m^2\right\}c_m^{(p)}\\
 &\quad+\frac{\eta_{AB}\omega_m^2}{2}
 \left(c_m^{(p-1)}+c_m^{(p+1)}\right)=0.
\end{aligned}
 \label{eq:app_recurrence}
\end{equation}
At marginal subharmonic stability, $s=0$; truncating this recurrence gives a finite linear system for the boundary of the first-order Mathieu tongue.

The finite-amplitude calculations instead use the period-dependent coefficient in Eq.~\eqref{eq:app_h_full}. Two linearly independent solutions of Eq.~\eqref{eq:app_Hill}, initially $(\zeta_m,\dot\zeta_m)=(1,0)$ and $(0,1)$, are integrated over $T_D$ to form the monodromy matrix $M_m$. With multipliers $\lambda_{m,+}$ and $\lambda_{m,-}$, the growing-mode amplitude exponent is
\begin{equation}
 \textcolor{black}{\Gamma_m}=\frac{1}{T_D}
 \max_{\nu=\pm}\ln|\lambda_{m,\nu}|.
 \label{eq:app_floquet_rate}
\end{equation}
Since Eq.~\eqref{eq:app_Hill} contains no first time derivative, $\det M_m=1$. The boundaries of the primary subharmonic tongue satisfy $\operatorname{tr}M_m=-2$. Repeating this calculation across driving frequency and relative modulation amplitude gives the tongues for $m=3$--$8$. The same static background parameters enter the first-order and finite-amplitude calculations.

\begin{figure}[tb]
 \centering
 \includegraphics[width=\columnwidth]{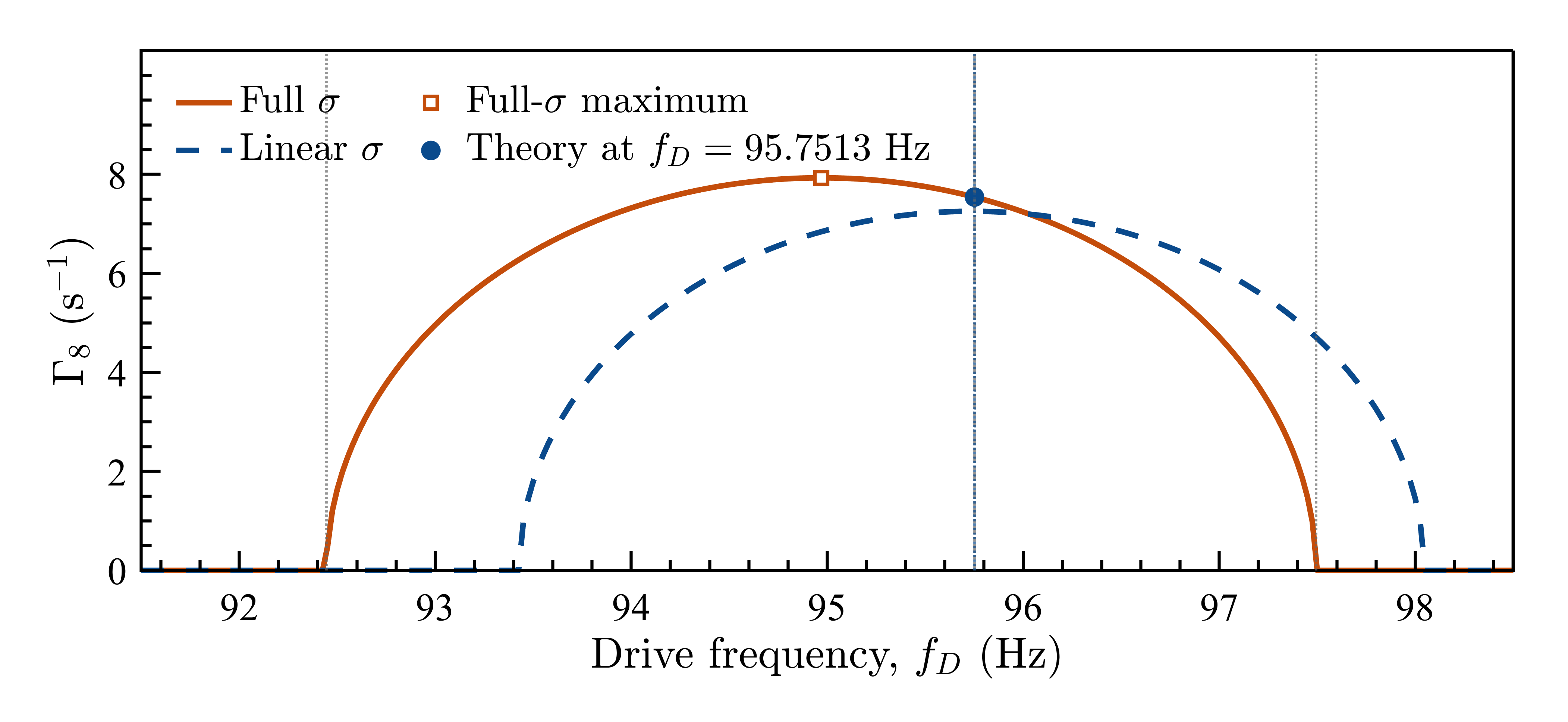}
 \caption{Theoretical Floquet amplitude growth rate of the eightfold mode as a function of driving frequency for $\epsilon_{AB}=168/426$, calculated with the unexpanded tension coefficient (solid) and the first-order response (dashed). The square marks the maximum of the unexpanded-tension result; the filled circle marks its value at $95.7513\,\mathrm{Hz}$.}
 \label{fig:app_floquet_rates}
\end{figure}

\section{Interface-mode analysis of the GP dynamics}
\label{app:GP_modes}

\subsection{Interface contour and angular amplitudes}
\label{app:GP_extraction}

At each time, the common internal boundary is represented by its polar radius $r(\theta,t)$. It is located at the crossing of the radial density profiles of the two components along rays extending from the center, selecting the internal boundary between component $B$ and component $A$ rather than the outer edge of the latter. The angularly averaged radius and the Fourier coefficients of its deformation are
\begin{align}
 R(t)&=\frac{1}{2\pi}\int_0^{2\pi}r(\theta,t)\,d\theta,
 \label{eq:app_mean_radius}\\
 q_m(t)&=\frac{1}{2\pi}\int_0^{2\pi}
 [r(\theta,t)-R(t)]e^{-im\theta}\,d\theta.
 \label{eq:app_q}
\end{align}
For a real interface radius, the contribution of a single $m$th harmonic is $2\operatorname{Re}[q_m(t)e^{im\theta}]$. Its nonnegative geometric amplitude is
\begin{equation}
 A_m(t)=2|q_m(t)|,
 \label{eq:app_A_m}
\end{equation}
which is distinct from the modulation amplitude $\Delta a_{AB}$. The phase-sensitive eightfold coordinate is obtained using a fixed initial phase $\phi_0=\arg q_8(0)$,
\begin{equation}
 X_8(t)=2\operatorname{Re}[q_8(t)e^{-i\phi_0}].
 \label{eq:app_signed_8}
\end{equation}
The coordinate $X_8$ retains the sign change of the standing deformation, whereas $A_8$ follows its magnitude.

\subsection{Period-averaged amplification}
\label{app:GP_average}

The slowly varying amplitude is characterized by its mean over one driving period,
\begin{equation}
 \overline A_m(t)=\frac{1}{T_D}
 \int_{t-T_D/2}^{t+T_D/2}A_m(t')\,dt',
 \label{eq:app_period_average}
\end{equation}
for times at which the complete averaging interval lies in the trajectory. This identical averaging prescription is applied to the driven and undriven trajectories in Fig.~\ref{fig:amplification}(a). For the interval $200$--$260\,\mathrm{ms}$, a linear fit to $\ln\overline A_8(t)$ gives
\begin{equation}
 \ln\overline A_8(t)\simeq b+\Gamma_{\mathrm{eff}}t,
 \qquad \Gamma_{\mathrm{eff}}\simeq13.9\,\mathrm{s^{-1}},
 \label{eq:app_GPeffective}
\end{equation}
where $b$ is a constant fit intercept. The inferred amplitude-doubling time is $\ln 2/\Gamma_{\mathrm{eff}}\simeq50\,\mathrm{ms}$. The fit characterizes a local, finite-time stage of the GP evolution, during which $A_8/R$ remains below $1\%$. It describes finite-time amplitude growth rather than a single asymptotic exponent for the full GP trajectory.

\begin{figure}[tb]
 \centering
 \includegraphics[width=\columnwidth]{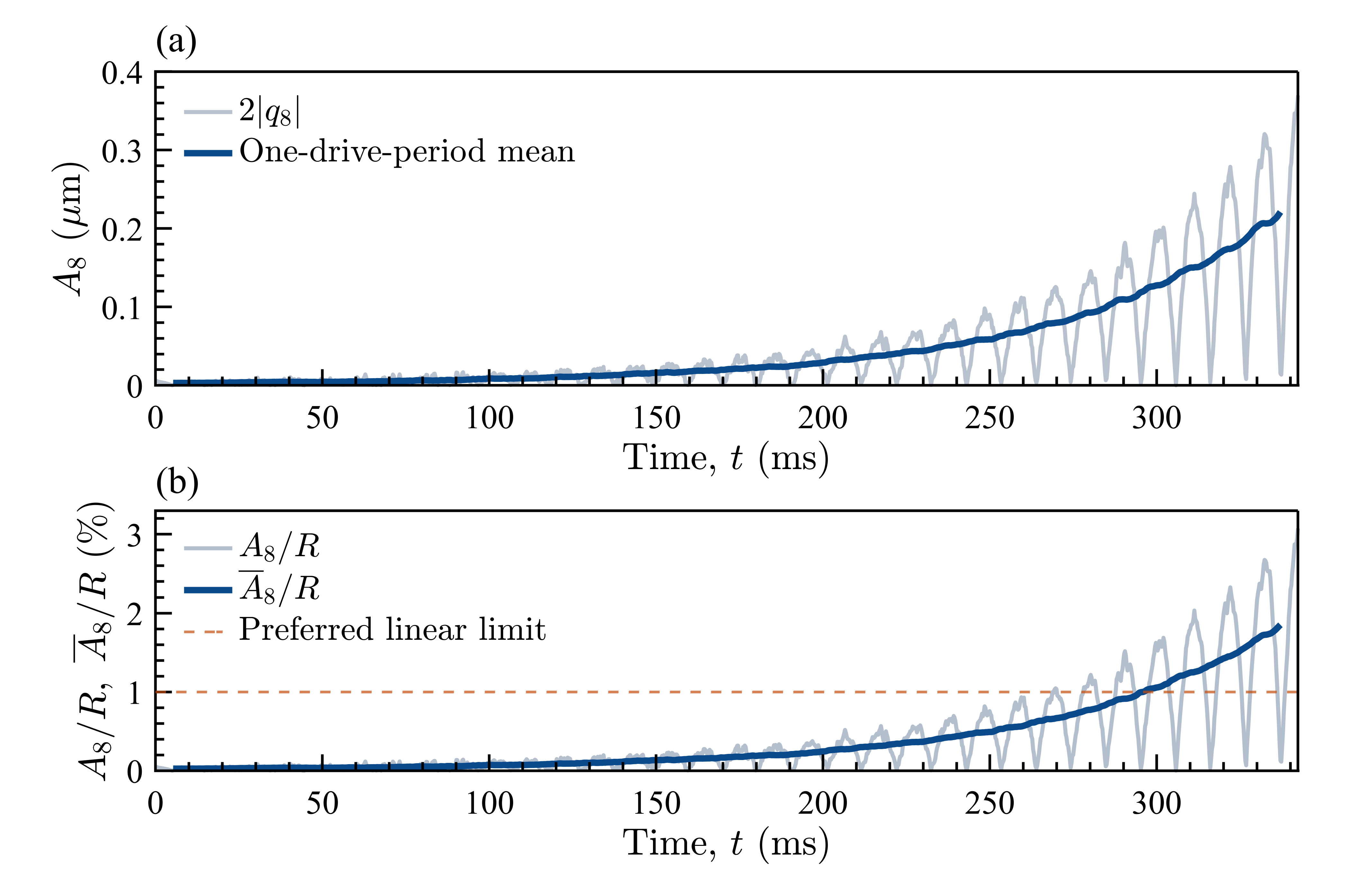}
 \caption{Raw and one-drive-period-averaged eightfold interface amplitudes in the driven GP trajectory. (a) Geometric amplitude. (b) Displacement relative to the instantaneous mean radius. The dashed $1\%$ line is the specified small-deformation reference for the local fit, not a measured onset of nonlinearity.}
 \label{fig:app_amplitude}
\end{figure}

The modal composition is characterized by the average squared geometric amplitudes, $\overline{A_m^2}$, obtained using the same period average as in Eq.~\eqref{eq:app_period_average}. For the analyzed modes $m=1,\ldots,32$, the fraction carried by the eighth component is
\begin{equation}
 S_8(t)=\frac{\overline{A_8^2}(t)}
 {\displaystyle\sum_{m=1}^{32}\overline{A_m^2}(t)}.
 \label{eq:app_modal_fraction}
\end{equation}
This quantity measures the angular composition of the interface displacement, rather than a fraction of the total GP energy. During $200$--$260\,\mathrm{ms}$, the eighth component contributes at least $82.9\%$ of the retained squared-amplitude sum and its period-averaged amplitude is at least $3.16$ times the largest competing component. These measurements describe the preferential amplification of an initially prepared eightfold deformation.

\subsection{Subharmonic phase and radial response}
\label{app:GP_phase}

The Fourier amplitude spectrum is computed from the mean-subtracted signed coordinate $X_8(t)$ using a Hann window and normalized by its maximum. The free-oscillation record spans approximately $400\,\mathrm{ms}$ and the driven record extends to $342.302\,\mathrm{ms}$. Their finite durations limit the resolution of the frequency estimates. The prominent driven spectral feature lies near $f_D/2$.

\begin{figure}[!htb]
    \centering
    \includegraphics[width=0.90\columnwidth]
    {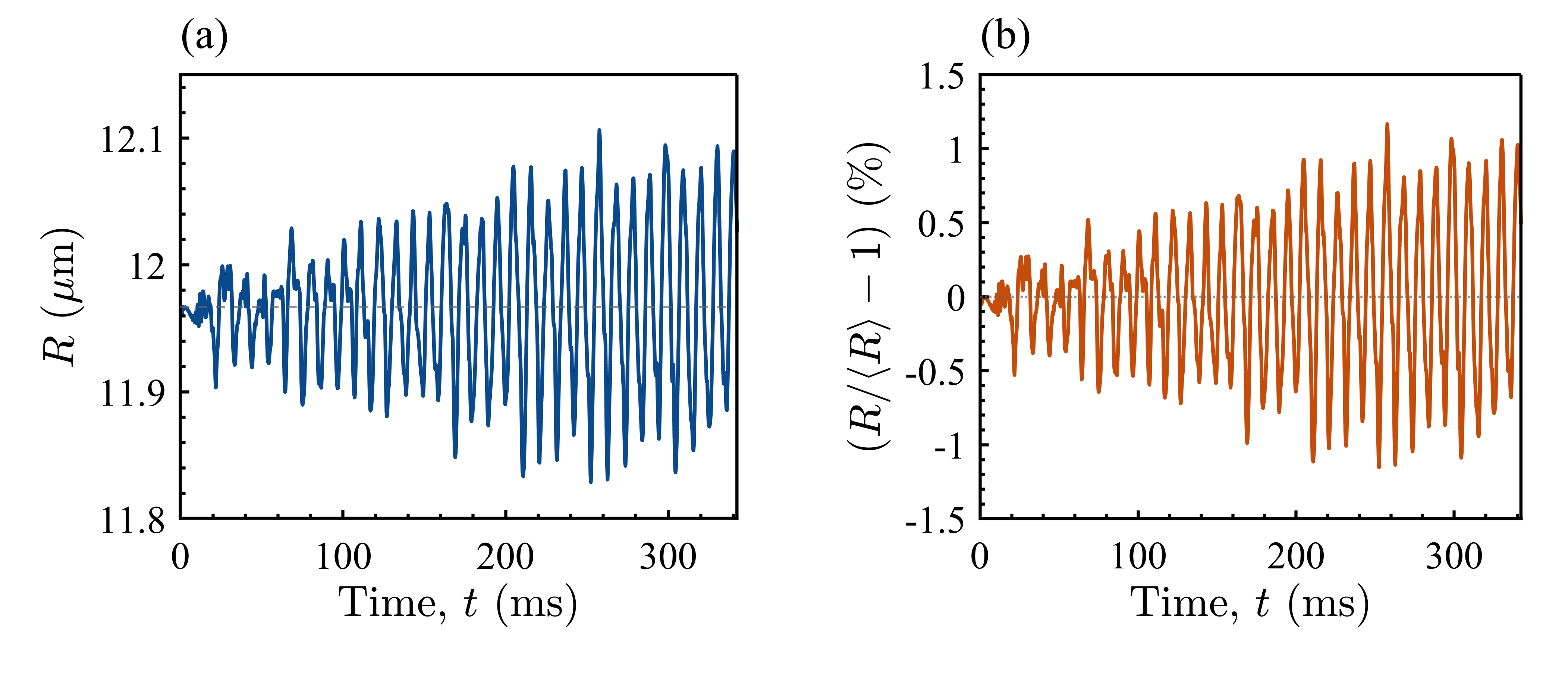}
    \caption{Radial motion of the common interface during the driven GP evolution.
    (a) Angularly averaged interface radius.
    (b) Relative variation of the interface radius about its time average.}
    \label{fig:app_radial}
\end{figure}

The phase recurrence can be quantified directly from the complex coefficient $q_8(t)$. For a shift by $s$ driving periods, the expected-sign-normalized inner product is
\begin{equation}
 \begin{aligned}
 C_s&=(-1)^s\operatorname{Re}
 \frac{\langle q_8,q_{8,s}\rangle}
 {\|q_8\|\,\|q_{8,s}\|},\\
 q_{8,s}(t)&=q_8(t+sT_D),\qquad s=1,2.
 \end{aligned}
 \label{eq:app_Cs}
\end{equation}
The inner product denotes $\langle u,v\rangle=\sum_t u^*(t)v(t)$ and the associated norm is $\|u\|^2=\sum_t|u(t)|^2$, with both series restricted to the same eligible reference times for each shift. Both times of every pair lie in the recorded trajectory; the shifted coefficient is evaluated by interpolation of the saved complex signal. The results $C_1=0.997487$ and $C_2=0.997053$ support an approximately $\pi$ phase change after $T_D$ and phase recovery after $2T_D$. The normalized inner product measures correspondence between the complex signals without requiring equal amplitudes at shifted times; it remains sensitive to changes in their relative amplitude envelopes.

\textcolor{black}{As shown in Fig.~\ref{fig:app_radial},} the angular mean $R(t)$ also oscillates during the driven GP evolution. Its peak-to-peak variation relative to its temporal mean is about $2.32\%$. The radial motion is an additional degree of freedom retained in the coupled GP equations and absent from the fixed-background capillary equation.


\end{document}